\documentclass[aps, prx, superscriptaddress, twocolumn, longbibliography, tightenlines, showpacs, floatfix, pdftex]{revtex4-1}

\usepackage{amsmath, amsfonts, amssymb}
\usepackage{bm,bbm}
\usepackage{mathtools}

\usepackage{afterpackage}
\usepackage{hyperref}
\hypersetup{
    colorlinks,
    linkcolor=blue,
    citecolor=blue,
    filecolor=blue,
    urlcolor=blue
}
\usepackage{multirow}
\usepackage{tikz} 
\usepackage[normalem]{ulem}
\usepackage{wasysym}

\usepackage{float}
\usepackage{graphicx}
\usepackage[caption=false]{subfig} 
\usepackage{array, booktabs}
	\newcolumntype{P}[1]{>{\centering\arraybackslash}p{#1}} 

\newcommand{\corr}[1]{\langle #1\rangle}

\newcommand{\bds}[1]{\boldsymbol{#1}}

\newcommand{\abs}[1]{\vert #1\vert}

\newcolumntype{C}{>{$}c<{$}}
\AtBeginDocument{
\heavyrulewidth=.08em
\lightrulewidth=.05em
\cmidrulewidth=.03em
\belowrulesep=.65ex
\belowbottomsep=0pt
\aboverulesep=.4ex
\abovetopsep=0pt
\cmidrulesep=\doublerulesep
\cmidrulekern=.5em
\defaultaddspace=.5em
}

\begin{document}

\title{Revealing the Phase Diagram of Kitaev Materials by Machine Learning: \\ Cooperation and Competition between Spin Liquids}
\date{\today}

\author{Ke Liu}
\email{ke.liu@lmu.de}
\affiliation{Arnold Sommerfeld Center for Theoretical Physics, University of Munich, Theresienstr. 37, 80333 M\"unchen, Germany}
\affiliation{Munich Center for Quantum Science and Technology (MCQST), Schellingstr. 4, 80799 M\"unchen, Germany}

\author{Nicolas Sadoune}
\affiliation{Arnold Sommerfeld Center for Theoretical Physics, University of Munich, Theresienstr. 37, 80333 M\"unchen, Germany}
\affiliation{Munich Center for Quantum Science and Technology (MCQST), Schellingstr. 4, 80799 M\"unchen, Germany}

\author{Nihal Rao}
\affiliation{Arnold Sommerfeld Center for Theoretical Physics, University of Munich, Theresienstr. 37, 80333 M\"unchen, Germany}
\affiliation{Munich Center for Quantum Science and Technology (MCQST), Schellingstr. 4, 80799 M\"unchen, Germany}

\author{Jonas Greitemann}
\affiliation{Arnold Sommerfeld Center for Theoretical Physics, University of Munich, Theresienstr. 37, 80333 M\"unchen, Germany}
\affiliation{Munich Center for Quantum Science and Technology (MCQST), Schellingstr. 4, 80799 M\"unchen, Germany}

\author{Lode Pollet}
\affiliation{Arnold Sommerfeld Center for Theoretical Physics, University of Munich, Theresienstr. 37, 80333 M\"unchen, Germany}
\affiliation{Munich Center for Quantum Science and Technology (MCQST), Schellingstr. 4, 80799 M\"unchen, Germany}
\affiliation{Wilczek Quantum Center, School of Physics and Astronomy, Shanghai Jiao Tong University, Shanghai 200240, China}

\begin{abstract}
Kitaev materials are promising materials for hosting quantum spin liquids and investigating the interplay of topological and symmetry-breaking phases.
We use an unsupervised and interpretable machine-learning method, the tensorial-kernel support vector machine, to study the honeycomb Kitaev-$\Gamma$ model in a magnetic field.
Our machine learns the global classical phase diagram and the associated analytical order parameters, including several distinct spin liquids, two exotic $S_3$ magnets, and two modulated $S_3 \times Z_3$ magnets.
We find that the extension of Kitaev spin liquids and a field-induced suppression of magnetic order already occur in the large-$S$ limit, implying that critical parts of the physics of Kitaev materials can be understood at the classical level.
Moreover, the two $S_3 \times Z_3$ orders are induced by competition between Kitaev and $\Gamma$ spin liquids and feature a different type of spin-lattice entangled modulation, which requires a matrix description instead of scalar phase factors.
Our work provides a direct instance of a machine detecting new phases and paves the way towards the development of automated tools to explore unsolved problems in many-body physics.
\end{abstract}

\maketitle

\section{Introduction} \label{sec:intro}
Kitaev materials have attracted immense attention in the search for quantum Kitaev spin liquids (KSLs)~\cite{Kitaev06}.
These materials feature highly anisotropic magnetic interactions, a necessary ingredient to realize the Kitaev model, and are found in Mott insulators with strong spin-orbit coupling~\cite{Jackeli09, Chaloupka10, Takagi19, Winter17b}.
Experimental signatures of the half-quantized thermal Hall effect, a key characteristic of spin-$1/2$ KSLs, in $\alpha$-$\mathrm{RuCl}_3$~\cite{Kasahara18, Yokoi20}, and the absence of noticeable magnetic orders in ${\rm H_3LiIr_2O_6}$~\cite{Kitagawa18} and ${\rm Cu_2IrO_3}$~\cite{Takahashi19} demonstrate that these materials are considered among the most prominent candidates for hosting spin liquids.
Theoretical studies have put forward an even greater variety of spin liquids and other exotic states~\cite{Song16, Kimchi11, Singh12, Price12, Li15, Sears15, Janssen16, Janssen17, Jiang19, Chern20, Gordon19, Wang19, Lee20, Gohlke18, Gohlke20,  Osorio14, Gohlke17, Motome20, Zhu18, Hickey19, Hickey20, Zhu20, Dong20, Berke20, Khait20, Rousochatzakis15} and generalized the family of Kitaev materials to high-spin systems~\cite{Stavropoulos19, Xu20}.
Three-dimensional hyper- and stripy-honeycomb materials are also synthesized in iridates $\beta$-, $\gamma$-${\rm Li}_2{\rm IrO}_3$ and are and under active investigation~\cite{Takagi19, Modic14, Takayama15, Biffin14, Ruiz17}.
Nevertheless, this enormous progress goes hand in hand with many open questions.
The role of non-Kitaev interactions, which generically exist in real materials, is yet to be understood.
The microscopic model of prime candidate compounds including $\alpha$-$\mathrm{RuCl}_3$ and the nature of their low-temperature phases remain under debate~\cite{Kim15, Kim16, Winter16, Yadav16, Ran17,  Hou17, Winter17, Eichstaedt19, Sears20, Banerjee17, Banerjee16, Koitzsch17,  Majumder15, Johnson15,  Cao16,  Lampen18, Balz19, Gass20, Wang17, Banerjee18, Wolter17, Lampen18b, Laurell20, Maksimov20, Bachus20}.
 Moreover,  conceptual understanding beyond the exactly solvable Kitaev limit largely relies on mean-field and spin-wave methods~\cite{Rau14, Rau16, Chaloupka15, Rusnacko19, Janssen19, Okamoto13}, as different numerical calculations of the same model Hamiltonian predict phase diagrams that are qualitatively in conflict with each other~\cite{Jiang19, Wang19, Gordon19, Lee20, Chern20, Gohlke18, Gohlke20}.

A data-driven approach such as machine learning may open an alternate route to research in Kitaev materials.
In recent years, its potential in physics has begun to be realized~\cite{Carleo19, Carrasquilla20}. Successful applications include representing quantum wave functions~\cite{Carleo17}, learning order parameters~\cite{Ponte17, Wang16}, classifying phases~\cite{Carrasquilla17, Nieuwenburg17}, designing algorithms~\cite{Liao19, LiuJW17}, analyzing experiments~\cite{Nussinov16, Zhang19} and optimizing material searches~\cite{Schmidt19}. Most of these advances are focused on algorithmic developments and resolving known problems.
Instead, it remains \emph{very rare} that such techniques are applied to a hard, unsolved problem in physics and provide new insights.

In this article, we employ our recently developed tensorial kernel support vector machine (TK-SVM)~\cite{Greitemann19, Liu19, Greitemann19b} to learn the global phase diagram of the honeycomb Kitaev-$\Gamma$ model under a $[111]$ field, which remains unsettled even in the (semi-)classical large-$S$ case.
The symmetric off-diagonal $\Gamma$ term is a typical non-Kitaev exchange present in real compounds and can originate from the direct overlap of $d$ orbitals and intermediate $d$-$p$ hopping~\cite{Rau14, Winter17}.
In particular, in $\alpha$-$\mathrm{RuCl}_3$ this exchange is believed to be comparable to the Kitaev interaction~\cite{Ran17, Kim16, Yadav16, Winter16, Wang17}.
Furthermore, it leads  to macroscopic degeneracies and classical spin liquids~\cite{Rousochatzakis17}.

We determine the global classical phase diagram of the $K$-$\Gamma$-$h$ model in a completely \emph{unsupervised} fashion.
The strong \emph{interpretability} of TK-SVM further allows us to achieve an analytical characterization of all phases. 
We hence provide a direct instance of a machine identifying new phases of matter in strongly-correlated condensed matter physics and show that the \emph{competition} and \emph{cooperation} between Kitaev and $\Gamma$ spin liquids are key in understanding the emergence of orders in the $K$-$\Gamma$ model.
We summarize our main findings below.

First, KSLs can survive non-Kitaev interactions in the large-$S$ limit.
The classical phase diagram shows remarkable similarities to its quantum counterpart in the subregion intensively investigated for spin-$1/2$ systems, including a field-induced suppression of magnetic order.
Second, the explicit emergent local constraints for classical $\Gamma$ spin liquids ($\Gamma$SLs) are found, and their local transformations are formulated.
Third, cooperation and competition between Kitaev and $\Gamma$ constraints lead to two $S_3$ orders and two $S_3 \times Z_3$ orders.
The latter features a different spin-lattice entangled modulation and may be realized by materials governed by strong Kitaev and $\Gamma$ interactions.

This article is organized as follows. In Section~\ref{sec:model_method} we define  the $K$-$\Gamma$-$h$ Hamiltonian and explain the essential ingredients of TK-SVM.
Section~\ref{sec:phase_diagram} is devoted to an overview of the machine-learned phase diagram.
Section~\ref{sec:sls} discusses the emergent local constraints of classical Kitaev and $\Gamma$ spin liquids and their local $Z_2$ symmetries.
The exotic $S_3$ and $S_3 \times Z_3$ orders are elaborated in Section~\ref{sec:orders}.
We conclude with an outlook in Section~\ref{sec:summary}.

\section{Model and method} \label{sec:model_method}
We subject the honeycomb Kitaev-$\Gamma$ model in a uniform $[111]$ field to the analysis of TK-SVM.
The spins will be treated as classical $O(3)$ vectors to achieve a  large system size which is important to capture competing orders induced by the $\Gamma$ interaction.

\emph{Hamiltonian.}
The $K$-$\Gamma$-$h$ Hamiltonian is defined as
\begin{align} \label{eq:Hamiltonian}
    H = \sum_{\corr{ij}_\gamma} \big[ K S_i^\gamma S_j^\gamma + \Gamma (S_i^\alpha S_j^\beta + S_i^\beta S_j^\alpha) \big] - \sum_i \vec{h} \cdot \vec{S}_i,
\end{align}
where $K$ and $\Gamma$ denote the strength of Kitaev and off-diagonal interactions, respectively;
$\gamma \in \{x, y, z \}$ labels the three different nearest-neighbor (NN) bonds $\corr{ij}_\gamma$;
$\alpha, \beta, \gamma$ are mutually orthogonal; and
$\vec{h} = h (1 \ 1 \ 1) / \sqrt{3}$.
We parametrize the interactions as
$K = \sin{\theta}$, $\Gamma = \cos{\theta}$, with $\theta \in [0, 2\pi)$.
The region $\theta \geq 3\pi/2$ corresponds to parameters of $4d$/$5d$ transition metals with ferromagnetic (FM) $K$~\cite{Takagi19},
while $\theta \in [\pi/2, \pi)$ relates to $4f$-electron based systems with anti-ferromagnetic (AFM) $K$~\cite{Jang19}.

\begin{figure*}[t]
  \centering
  \includegraphics[width=0.92\textwidth]{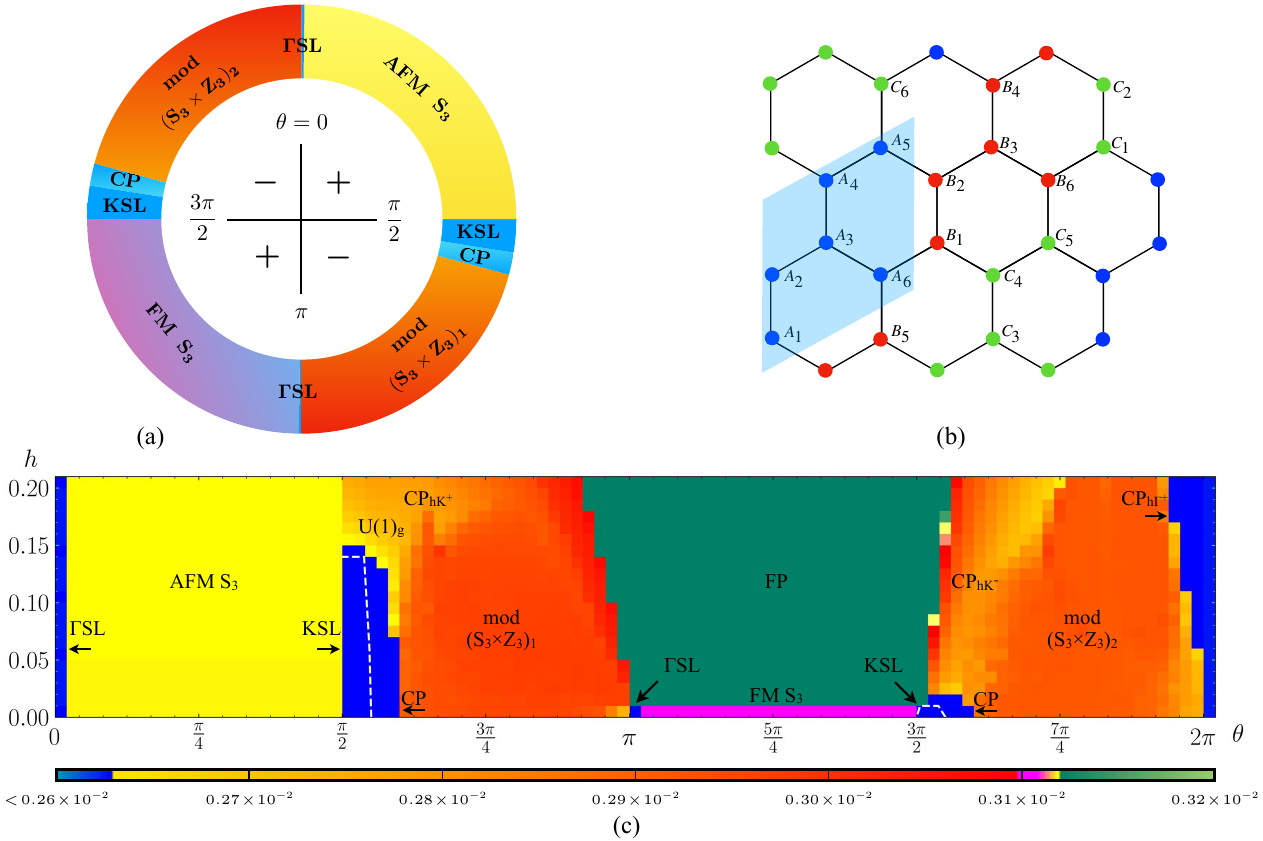}
  \caption{Machine-learned phase diagram for the honeycomb $K$-$\Gamma$ model in a $[111]$ magnetic field, with $K = \sin\theta, \Gamma=\cos\theta$ and at temperature $T = 10^{-3}\sqrt{K^2 + \Gamma^2}$.
  ({\bf a}) Circular representation of the $h = 0$ phase diagram as a function of angle $\theta$. Classical $\Gamma$ ($\Gamma$SLs) and Kitaev (KSLs) spin liquids reside in the limits $\theta \in \{0, \frac{\pi}{2}, \pi, \frac{3\pi}{2}\}$ [$(K, \Gamma) = (0, \pm 1), (\pm 1, 0)$].
  These special limits divide the phase diagram into two frustrated ($K\Gamma < 0$) and two unfrustrated ($K\Gamma > 0$) regions, labeled by ``$-$'' and ``$+$', respectively.
  While $\Gamma$SLs exist only in the two large $\Gamma$ limits, KSLs extend into the frustrated regions, until $\abs{\Gamma/K}_1 \sim 0.16$ ($\theta \sim 0.551 \pi, 1.551 \pi$).
From $\abs{\Gamma/K}_2 \sim 0.27$ ($\theta \sim 0.584 \pi, 1.584 \pi$), two modulated $S_3 \times Z_3$ orders will be stabilized owing to competition between a KSL and a $\Gamma$SL.
  These orders have a highly exquisite magnetic structure featuring spin-lattice entangled modulation.
  In the windows between KSLs and the modulated $S_3 \times Z_3$ orders, there are two non-Kitaev correlated paramagnets (CPs).
  The two unfrustrated regions respectively host a ferromagnetic (FM) and an antiferromagnetic (AFM) $S_3$ order, induced by cooperation between KSLs and $\Gamma$SLs.
  The $h=0$ phase diagram is symmetric under $\theta \rightarrow \theta + \pi$ and a sublattice transformation (see Section~\ref{sec:model_method}).
  ({\bf b})~Magnetic cells of the $S_3$ and  modulated $S_3 \times Z_3$ orders. The shaded sites show a magnetic cell for the FM and AFM $S_3$ order, comprised of six spins.
  The modulated $S_3 \times Z_3$ orders consist of three distinct $S_3$ sectors (labeled by $A,B,C$) and in total eighteen sublattices (Section~\ref{sec:orders}).
  ({\bf c}) Finite $h$ phase diagram.
  The FM $S_3$ and the KSL ($\Gamma$SL) for $K = -1$ ($\Gamma = -1$) will be fully polarized (FP) once the $[111]$ field is applied.
  However, an antiferromagnetic $\Gamma$ extends the FM KSL to a small, but finite, $h \sim 0.01$.
  AFM $\Gamma$SL and AFM KSL are robust against external fields. The former persists until $h \geq 0.2$, while the latter is non-trivially polarized from $h \sim 0.14$ with global $U(1)$-symmetric correlations [$U(1)_g$].
  In the frustrated regions and intermediate fields, there are areas of different partially-polarized correlated paramagnets (${\rm CP_h}$s).
  In particular, in the sector of $K < 0, \Gamma > 0$, the ${\rm CP_{hK^-}}$ and ${\rm CP_{h \Gamma^+}}$ regimes erode the modulated $(S_3 \times Z_3)_2$ phase, as field-induced suppression of magnetic order takes hold.
  Each pixel in the phase diagram represents a $(\theta, h)$ point
   and is color-coded by the corresponding Fiedler vector entry.
  The sharp jumps in color coincide with the well-separated peaks in the distribution of Fiedler vector entries, while gentle gradients are indicative of crossovers; cf. Appendix~\ref{app:graph}.
  Dashed lines separate a spin liquid from a correlated paramagnet, based on susceptibility of the associated ground state constraint (GSC).
  The Fiedler vector and the GSCs are computed from rank-$1$ and rank-$2$ TK-SVM, respectively.
  See the text and Appendix~\ref{app:graph}, Appendix~\ref{app:measure} for details.}
  \label{fig:PD}
  \clearpage 
\end{figure*}

The Hamiltonian given by Eq.~\eqref{eq:Hamiltonian} features a global $C_6^R C_3^S$ symmetry which acts simultaneously on the real and spin space, where $C_6^R$ rotates the six spins on a hexagon (anti-)clockwise,
and $C_3^S$ (anti-)cyclically permutates $\{S^x, S^y, S^z\}$.
In the absence of magnetic fields, the Hamiltonian is also symmetric under a sublattice transformation by sending
 $K \rightarrow -K$, $\Gamma \rightarrow -\Gamma$, and meanwhile $S_i \rightarrow -S_i$ for either of the honeycomb sublattices.
This sublattice symmetry indicates equivalence between the $K$-$\Gamma$ model of FM and AFM Kitaev interaction, which is respected by the $h = 0$ phase diagram shown in Figure~\ref{fig:PD} (a) and the associated order parameters.

\emph{Machine learning.}
The TK-SVM is defined by the decision function
\begin{align}\label{eq:d(x)}
    d(\mathbf{x}) = \sum_{\mu\nu} C_{\mu\nu} \phi_\mu(\mathbf{x}) \phi_\nu(\mathbf{x}) - \rho.
\end{align}
Here, $\mathbf{x} = \{S_i^x, S_i^y, S_i^z | i = 1,2, \dots, N\}$ denotes a spin configuration of $N$ spins, which is the only required input. No prior knowledge of the phase diagram is required.

$\bds{\phi}(\mathbf{x})$ denotes a feature vector mapping $\mathbf{x}$ to an auxiliary feature space.
When orders are detected, they are encoded in the coefficient matrix $\mathbb{C} = \{ C_{\mu\nu} \}$.
The first term in $d(\mathbf{x})$ captures both the form and the magnitude of orders in the system, regardless of whether they are unconventional magnets, hidden nematics~\cite{Greitemann19, Liu19} or classical spin liquids~\cite{Greitemann19b}.
The extraction of analytical order parameters is straightforward in virtue of the strong interpretability of SVM (see Appendix~\ref{app:op} for details).

The second term $\rho$ in the decision function is a bias parameter and reflects an order-disorder hierarchy between two sample sets.
It detects whether samples in one training set are more ordered or disordered than those in the other set, hence allows one to infer if two states belong to the same phase~\cite{Liu19, Greitemann19b}.
This property of the $\rho$ parameter leads to a graph analysis.
By treating points in the physical parameter space as vertices and assigning an edge to any two vertices, one can create a graph with the edge weights determined by $\rho$.
Computing the phase diagram is then realized by an unsupervised graph partitioning (see Appendix~\ref{app:graph}).

The concrete application of TK-SVM consists of several steps.
First, we collect samples from the parameter space of interest.
For the $K$-$\Gamma$-$h$ model, large-scale parallel-tempering Monte Carlo simulations~\cite{Hukushima96, BookBinder} are utilized to generate those configurations, with system sizes up to $N = 10,\!368$ spins ($72\times72$ honeycomb unit cells).
As major parts of the phase diagram are unknown, we distribute the phase points (almost) uniformly in the $\theta$-$h$ space.
In total, $M = 1,\!250$ distinct $(\theta, h)$-points at low temperature $T = 10^{-3} \sqrt{K^2 + \Gamma^2}$ are collected; each has $500$ sufficiently uncorrelated samples.
Then, we perform a SVM multi-classification on the sampled data.
From the obtained $\rho$'s, we build a graph of $M$ vertices and $M(M-1)/2$ edges and partition it by Fiedler's theory of spectral clustering~\cite{Fiedler73, Fiedler75}.
The outcome is the so-called Fiedler vector reflecting clustering of the graph, which plays the role of the phase diagram [see Figure~\ref{fig:PD} (c)].
In the next step, based on the learned phase diagram, we collect more samples (typically a few thousands) for each phase and perform a separate multiclassification.
The goal here is to learn the $C_{\mu\nu}$ matrices of high quality in order to extract analytical quantities. The dimension of this reduced classification problem depends on the number of phases (subgraphs).
Finally, we measure the learned quantities to validate that they are indeed the correct order parameters.

\section{Global view of the phase diagram} \label{sec:phase_diagram}

The $K$-$\Gamma$-$h$ model shows a rich phase diagram, including a variety of classical spin liquids and exotic magnetic orders.
In the vicinity of the ferromagnetic Kitaev limit  with $\Gamma \gtrsim 0$ (i.e. $\theta \gtrsim \frac{3\pi}{2}$), which has been intensively studied for spin-$1/2$ systems, the classical phase diagram shares a number of important features with the quantum counterpart.
We will focus here on the topology of the machine-learned phase diagram. The specific properties of each phase are analyzed in subsequent sections.

We first discuss the phase diagram at $h = 0$, depicted in Figure~\ref{fig:PD} (a).
In the absence of external fields the Hamiltonian Eq.~\eqref{eq:Hamiltonian} has four limits at
$(K, \Gamma) = (\pm 1, 0)$ and  $(0, \pm 1)$,
corresponding to two classical KSLs and two $\Gamma$SLs.
These particular limits divide the $K$-$\Gamma$ phase diagram into four regions.
When both the Kitaev and $\Gamma$ interactions are ferromagnetic or antiferromagnetic, the system is unfrustrated,
while when they are of different sign, the system stays highly frustrated.

In the two unfrustrated $K\Gamma > 0$ regions, when $K$ and $\Gamma$ are both finite, the system immediately changes from a spin liquid to a magnetic order, which is sometimes described as a $120^{\circ}$ state~\cite{Rau14, Chaloupka15}.
The explicit order parameter of the two phases corresponds to the symmetric group $S_3$, hence we refer to them as the FM  $S_3$ and AFM $S_3$ phase, respectively, to distinguish them from other types of $120^{\circ}$ states.
As we shall see in Section~\ref{sec:orders}, these two orders can be understood as the result of  \emph{cooperation} between the Kitaev and $\Gamma$ spin liquids.

The physics is profoundly different in the frustrated regions.
The two KSLs can extend to a finite value of $\Gamma$ for $K\Gamma < 0$.
There has been mounting evidence suggesting that quantum KSLs survive in some non-Kitaev interactions~\cite{Kasahara18, Yokoi20, Gohlke18, Gohlke20, Lee20, Wang19, Gordon19, Osorio14, Gohlke17}.
It is quite remarkable that such an extension already manifests itself in the classical large-$S$ limit.
Using the corresponding ground state constraint (GSC), we estimate $\abs{\Gamma/K} \sim 0.16$ (see Appendix~\ref{app:measure}).
This large extension may find its origin in  the large extensive ground-state degeneracy (exGSD) of classical KSLs.

 By contrast, the two classical $\Gamma$SLs are found to only exist in the limit $\Gamma = \pm 1$, as in these cases the exGSD is much smaller (Cf. Section~\ref{sec:sls}).

The majority of the frustrated regions are occupied by two exotic orders.
In the ferromagnetic $K$ sector, it has been recently proposed to accommodate incommensurate orders or disordered states by numerical studies based on small system sizes~\cite{Jiang19, Gordon19, Chern20}.
However, by learning the explicit order parameter (Section~\ref{sec:orders}), our machine reveals that the order there, as well as its counterpart on the antiferromagnetic $K$ sector, have a more intriguing structure.
They possess threefolds of the magnetic structure discussed for the FM and AFM $S_3$ phase, leading to eighteen sublattices.
The three $S_3$ sectors mutually cancel via a novel modulation, and we henceforth refer to them as modulated $S_3 \times Z_3$ phase.
We also find out that \emph{competition} between a Kitaev and a $\Gamma$ spin liquid induces these orders.

Between each modulated $S_3 \times Z_3$ phase and the corresponding KSL, there is a window of another correlated disordered region.
It may be understood as a crossover between the two phases, as we are considering $O(3)$ spins at two dimensions and finite temperature.
We refer to such regions as correlated paramagnets (CP), which however may shrink in size in case the phase transitions get sharper.

When the $[111]$ magnetic field is turned on, the fate of each phase strongly depends on the sign of its interactions, as is shown in Figure~\ref{fig:PD} (c).
Those featuring only ferromagnetic interactions, including the FM $S_3$ phase, the FM Kitaev and $\Gamma$ spin liquids, immediately polarize.
However, the phases with one or both antiferromagnetic interactions are robust against finite $h$.
Specifically, the AFM KSL persists up to $h \sim 0.14$.
And before trivial polarization occurs at much stronger fields, there exists an intermediate region, dubbed $U(1)_g$, where the magnetic field induces two different correlations with a \emph{global} $U(1)$ symmetry (Section~\ref{sec:sls}).
Interestingly, this region appears to coincide with a gapless spin liquid phase recently proposed for quantum  spin-$1/2$ and spin-$1$ systems~\cite{Motome20, Zhu18, Hickey19, Hickey20, Zhu20}.

The frustrated $K\Gamma < 0$ regions are again richest in physics.
The FM KSL extends to a small, but finite, field $h \sim 0.01$ thanks to an antiferromagnetic $\Gamma$, while the AFM KSL extends over a much greater area.
At intermediate $h$, there are disordered regions separating a $S_3 \times Z_3$ phase from a spin liquid or a trivially polarized state.
We refer to them as partially-polarized correlated paramagnets (${\rm CP_h}$s) to distinguish them from the parent spin liquid.
In particular, the ${\rm CP_{hK^-}}$ and ${\rm CP_{h \Gamma^+}}$ regimes erode the modulated $(S_3 \times Z_3)_2$ phase (see Appendix~\ref{app:S3Z3}).
It is worth mentioning that a field-induced unconventional paramagnet has also recently been proposed for quantum spin-$1/2$ in the ${\rm CP_{hK^-}}$ region~\cite{Lee20, Gohlke20}.
These common features indicate that some critical properties of Kitaev materials, for those where Kitaev and $\Gamma$ interactions play a significant role, may already be understood at the classical level.

Before delving deeper into each phase, we comment on the distinctions between the graph partitioning in TK-SVM and traditional approaches of computing phase diagrams.
In learning the finite-$h$ phase diagram Figure\mbox{~\ref{fig:PD}(c)}, we did not use particular order parameters, nor any form of supervision.
Instead, $M(M-1)/2 = 780,\!625$ distinct decision functions are implicitly utilized; each serves as a classifier between two $(\theta, h)$ points (see Appendix~\ref{app:graph}).
Moreover, all phases are identified at once, rather than individually scanning each phase boundaries.
These make TK-SVM an especially efficient framework to explore phase diagrams with complex topology and unknown order parameters.

\section{Emergent Local constraints} \label{sec:sls}

A common feature of classical spin liquids is the existence of a non-trivial GSC which is an emergent local quantity that defines the ground-state manifold and controls low-lying excitations.
A system can be considered as a classical spin liquid if it breaks no orientation symmetry, and meanwhile its GSC has a local symmetry.
We now discuss the GSCs learned by TK-SVM for the classical Kitaev and $\Gamma$ spin liquids.

Our machine learns a distinct constraint for each spin liquid in the phase diagram Figure~\ref{fig:PD}.
These constraints can be expressed in terms of quadratic correlations on a hexagon.
We classify six types of such correlations at $h = 0$ and another two field-induced correlations for the AFM KSL, as tabulated in Table~\ref{tab:containts}.

\begin{table}[!tp]
\renewcommand{\arraystretch}{1.618}
\begin{tabular}{p{0ex}>{} p{5.4cm}*{2} {p{1.2cm} p{1.2cm}}}
 &        & \multicolumn{2}{c}{Symmetry} \\
  \cmidrule(r){3-4}

 &     Correlations       & Global & Local  \\
 \midrule
& $G_1 = \sum\limits_{\corr{ij} \in \varhexagon} S_i^\gamma S_j^\gamma$ & $C_6^R C_3^S$ & $Z_2$   \\
& $G_2 = \sum\limits_{\corr{ij} \in \varhexagon} \sum\limits_{\alpha\beta} \abs{\varepsilon_{\alpha\beta\gamma}} S_i^\alpha S_j^\beta $ & $C_6^R C_3^S$ & $Z_2$ \\
& $G_3 = \sum\limits_{[ij] \in \varhexagon} S_i^{\gamma_2} S_j^{\gamma_1}$ & $C_6^R C_3^S$ & $Z_2$  \\
& $G_4 = \sum\limits_{[ij] \in \varhexagon} \abs{\varepsilon_{\alpha\gamma_1 \gamma_2}}
        (S^{\gamma_1}_i S_j^\alpha  + S_i^\alpha S_j^{\gamma_2})$ & $C_6^R C_3^S$ \\
& $G_5 = \sum\limits_{(ij) \in \varhexagon} S_i^c S_j^c$ & $C_6^R C_3^S$ & $Z_2$ \\
& $G_6 = \sum\limits_{(ij) \in \varhexagon} \sum\limits_{ab} \abs{\varepsilon_{ab c}} S_i^a S_j^b $ & $C_6^R C_3^S$ \\
 \hline
& $G_1^h = \sum\limits_{\corr{ij} \in \varhexagon} \sum\limits_{\alpha\beta} S_i^\alpha S_j^\beta$ & $U(1)$
    \\
& $G_2^h = \sum\limits_{(ij) \in \varhexagon} \sum\limits_{ab} S_i^a S_j^b$ & $U(1)$ \\
 \bottomrule
\end{tabular}
\caption{Quadratic correlations classified by rank-$2$ TK-SVM.
$G_{\rm KSL} = \frac{1}{2}\corr{G_1}_{\varhexagon} = \pm 1$ and $\corr{G_{k \neq 1}}_{\varhexagon} = 0$ define the grounds states of FM and AFM KSLs, respectively.
$G_{\rm \Gamma SL} =  \frac{1}{7} \langle G_2 \pm G_3 +G5 \rangle_{\varhexagon} = \pm 1$ and vanishing $G_1, G_4,  G_6$ define the ground states of FM and AFM $\Gamma$SLs.
For the two $S_3$ orders, all $G_k$ contribute with an equal weight.
No stable ground-state constraints are found in the modulated $S_3 \times Z_3$ phases and those correlated paramagnetic regions.
All $G_k$ preserve the global $C^R_6 C^S_3$ symmetry of the $K$-$\Gamma$-$h$ Hamiltonian Eq.~\eqref{eq:Hamiltonian}.
$G_1$ ($G_2, G_3, G_5$) is locally $Z_2$ invariant on a bond (hexagon).
$G^h_1, G^h_2$ are field-induced correlations for the AFM Kitaev model with a global $U(1)$ symmetry.
See texts for details and Figure~\ref{fig:hexagon} for an illustration of the convention.
}\label{tab:containts}
\end{table}

For KSLs, we reproduce the GSCs previously obtained by a Jordan-Wigner construction~\cite{Baskaran08},
\begin{align} \label{eq:KSL}
        G_{\rm KSL} =  \frac{1}{2} \corr{G_1}_{\varhexagon} = \pm 1, \quad
        \corr{G_{k \neq 1}}_{\varhexagon} = 0,
\end{align}
where $``\pm"$ corresponds to the FM and AFM interaction, respectively (the same convention used below);
$\corr{\dots}_{\varhexagon}$ denotes the thermal average over hexagons.
As discussed in Refs.~\onlinecite{Baskaran08, Sela14}, these constraints impose degenerate dimer coverings on a honeycomb lattice, which are precisely the ground states of classical KSLs.

In  case of classical $\Gamma$SLs, our machine identifies two constraints,
\begin{align} \label{eq:GSL}
        G_{\rm \Gamma SL} &=  \frac{1}{7} \langle G_2 \pm G_3 +G_5 \rangle_{\varhexagon} = \pm 1, \nonumber \\
        \corr{G_1}_{\varhexagon} &= \corr{G_4}_{\varhexagon} = \corr{G_6}_{\varhexagon} = 0,
\end{align}
which directly differentiate between the FM and AFM case, and satisfying them will naturally lead to the ground-state flux pattern $W \sim \{1, 0, 0\}$ for every three hexagon plaquettes~\cite{Rousochatzakis17, Saha19}, where $W = S_1^x S_2^z S_3^y S_4^x S_5^z S_6^y$.

Aside from manifesting ground state configurations, knowing the explicit GSC will make clear the symmetry properties and the extensive degeneracy of a spin liquid.
The above Kitaev and $\Gamma$ constraints preserve the global $C^{R}_6 C^{S}_3$ symmetry of the Hamiltonian Eq.~\eqref{eq:Hamiltonian}, and more importantly, possess a different \emph{local} $Z_2$ symmetry, representing distinct classical $Z_2$ spin liquids.

The Kitaev constraints Eq.~\eqref{eq:KSL} are invariant by locally flipping the $\gamma$ component of a NN bond $\corr{ij}_{\gamma}$,
\begin{align}\label{eq:sym_KSL}
    S_i^{\gamma} \rightarrow - S_i^{\gamma}, \ S_j^{\gamma} \rightarrow - S_j^{\gamma}, \ i,j \in \corr{ij}_\gamma \in G_1.
\end{align}
For a given dimer covering configuration, this will give rise to $(2^3)^{1/3}$ redundant degrees of freedom on each hexagon.
Together with the $1.381^{N/2}$ dimer coverings on a honeycomb lattice~\cite{Wu06,Baxter70,Kasteleyn63}, it enumerates $1.662^N$ \emph{extensively} degenerate ground states~\cite{Baskaran08}, resulting in a residual entropy $\frac{S}{N} \approx 0.508$ at zero temperature.

\begin{figure}[t]
  \centering
  \includegraphics[width=0.45\textwidth]{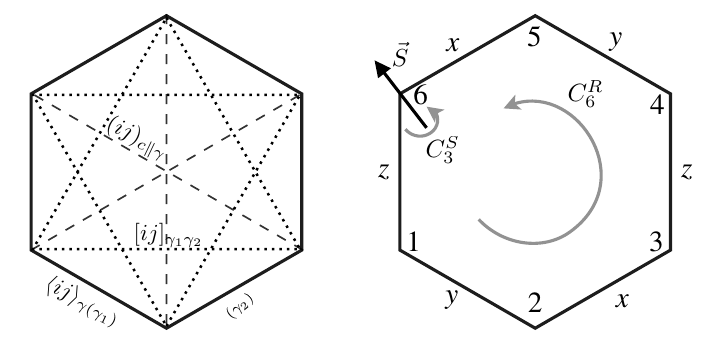}
  \caption{Convention of the quadratic correlations in Table~\ref{tab:containts}.
  $\corr{.}$, $[.]$ and $(.)$ denote the first, second and third nearest-neighbor (NN) bonds, respectively.
  $\gamma = x, y, z$ label the type of a NN bond.
  $\gamma_1, \gamma_2$ correspond to the two connecting NN bonds.
  $c = \gamma$ is determined by the parallel NN bond.
  $\alpha, \beta, \gamma$ ($a, b, c$) are mutually orthogonal.
  $C^R_6$ is a symmetry that rotates the six spins on a hexagon (anti-)clockwise.
  $C_3^S$ denote (anti-)cyclic permutations of the three spin components.
  }\label{fig:hexagon}
\end{figure}

The local invariance of the $\Gamma$SL constraints Eq.~\eqref{eq:GSL} takes a different form, defined on a hexagon,
\begin{gather}
    S_i^{\alpha} \rightarrow - S_i^{\alpha}, \ \ S_j^{\beta} \rightarrow - S_j^{\beta},
    \ \forall \corr{ij}_{\alpha,\beta \perp \gamma} \in G_2, \nonumber \\
    S_i^{\gamma_2} \rightarrow - S_i^{\gamma_2}, \ S_j^{\gamma_1} \rightarrow - S_j^{\gamma_1},
    \ \forall [ij]_{\gamma_1 \gamma_2} \in G_3, \nonumber \\
    S_i^{c} \rightarrow - S_i^{c}, \ S_j^{c} \rightarrow - S_j^{c},
     \ \forall (ij)_{c \parallel \gamma} \in G_5. \label{eq:sym_GSL}
\end{gather}
Here, $\alpha, \beta$ are the components normal to $\gamma$;
``$[.]$'' denotes the second nearest-neighbor bonds with $\gamma_1, \gamma_2$ corresponding to the two connecting NN bonds;
``$(.)$'' denotes the third nearest-neighbor bonds, and $c$ equals the $\gamma$ on a parallel NN bond; as depicted in Figure~\ref{fig:hexagon}.
This symmetry is considerably involved but also evident once the explicit GSC is identified.

The corresponding exGSD can again be counted by the local redundancy on a hexagon, giving $2^{N/6} \approx 1.122^N$ with a residual entropy $\frac{S}{N} \approx 0.115$.
This degeneracy is exponentially less than that of KSLs. As a result, $\Gamma$SLs are more prone to fluctuations (see Figure~\ref{fig:PD} and~\ref{fig:measure}).

Furthermore, in addition to the constraints for ground states, in the $U(1)_g$ region in the phase diagram Figure~\ref{fig:PD} (c), we identify two field-induced quadratic correlations.
The two correlations, denoted as $G^h_1$ and $G^h_2$ in Table~\ref{tab:containts}, are invariant under global rotations about the direction of the $\vec{h}_{\rm 111}$ fields.
From general symmetry principle, a continuous global symmetry will naturally support gapless modes.
Hence, aside from being different local observables in the classical AFM Kitaev model, they may also shine light on the nature of the corresponding gapless quantum spin liquid~\cite{Motome20, Zhu18, Hickey19, Hickey20, Zhu20}.

Note that the GSCs and other quadratic correlations learned by TK-SVM are not limited to classical spins.
Their formalism holds for general spin-$S$ and can be directly measured in the quantum $K$-$\Gamma$ model.
Comparing to other quantities (such as plaquette fluxes, Wilson/Polyakov loops, and spin structure factors), which may exhibit similar behaviors in different spin liquids, GSCs can be made unique to a ground-state manifold and hence may be more distinctive.
Moreover, their violation  provides a natural way to measure the breakdown of a spin liquid, which is what  we use to estimate the extension of KSLs (see Appendix~\ref{app:measure}).

\section{Cooperative and competitive constraint-induced ordering} \label{sec:orders}

\begin{figure}[tp!]
\centering
     \includegraphics[width=0.48\textwidth]{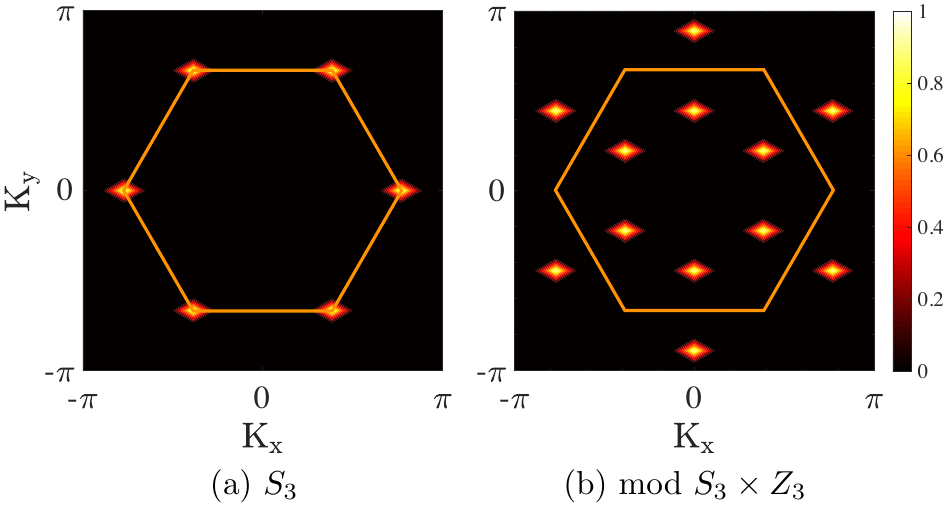}
  \caption{Static spin-structure factors (SSFs), 
  $S (\vec{K}) =  \big\langle \frac{1}{N} \sum_{ij} \vec{S}_i \cdot \vec{S}_j \, e^{i \vec{K} \cdot (\vec{r}_i - \vec{r}_j)}
  \big\rangle$, where $\vec{r}_i$ is the position of a spin at site $i$, and $\langle . \rangle$ denotes the ensemble average. 
  The two $S_3$ orders develop magnetic Bragg peaks at the $\mathbf{K}$ points of the honeycomb Brillouin zone (orange hexagon).
  The two $S_3 \times Z_3$ orders show Bragg peaks at $\frac{2}{3}\mathbf{M}$ points, owing to the larger magnetic cell.
  The length of nearest-neighbor bonds of the honeycomb lattice is set to unity.
  }\label{fig:SSF}
\end{figure}

\begin{figure}[t!]
  \centering
  \includegraphics[width=0.47\textwidth]{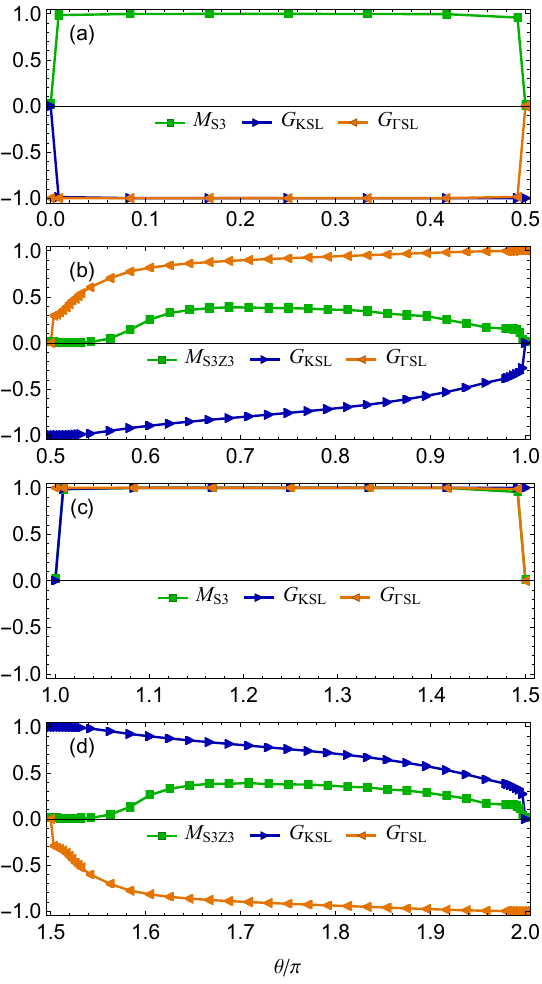}
  \caption{Measurements of the $S_3$ and modulated $S_3 \times Z_3$ magnetizations (green), and characteristic Kitaev (blue) and $\Gamma$ (orange) correlations, with $K = \sin\theta$, $\Gamma = \cos\theta$, $h = 0$, $T = 0.001$.
  $M = \big\langle\abs{\frac{1}{N_{\rm cell}}\sum_{\rm cell}\protect\overrightarrow{M}}\big\rangle$ measures the magnitude of the respective magnetization, where $\protect\overrightarrow{M}$ denotes the ordering moments in one magnetic cell, $\frac{1}{N_{\rm cell}}\sum_{\rm cell} (.)$ and $\langle . \rangle$ indicate the lattice and ensemble average, respectively.
  At the Kitaev ($\theta = \frac{\pi}{2}, \frac{3\pi}{2}$) and $\Gamma$ ($\theta = 0, \pi$) limits, either $G_{\rm KSL} = \pm1$ or $G_{\rm \Gamma SL} = \pm1$, satisfying the corresponding ground-state constraint.
  In the unfrustrated regions, $K\Gamma > 0$, Kitaev and $\Gamma$ correlations behave in an equal footing as $G_{\rm KSL} = G_{\rm \Gamma SL} = \pm 1$, and cooperatively induce the AFM~(a) or FM~(c) $S_3$ order.
  In the frustrated regions, $K\Gamma < 0$ [(b), (d)], $G_{\rm KSL}$ and $G_{\rm \Gamma SL}$ develop towards opposite directions.
  Though the system stays disordered near the Kitaev limits, from $\abs{\Gamma/K} \sim 0.27$ up to the large $\Gamma$ limits, the $S_3 \times Z_3$ orders are established owing to the competition between $G_{\rm KSL}$ and $G_{\rm \Gamma SL}$.
  }\label{fig:measure}
  \clearpage
\end{figure}

\begin{table*}[t!]
\centering
\renewcommand{\arraystretch}{2}
\newcolumntype{C}[1]{>{\centering\arraybackslash$}m{#1}<{$}}
\newlength{\mycolwd}
\newcommand\SmallMatrix[1]{\settowidth{\mycolwd}{$a-1$}\scalebox{0.618}{\renewcommand{\arraystretch}{0.85}\ensuremath{\left(\begin{array}{*{3}{@{}C{\mycolwd}@{}}} \displaystyle #1 \end{array}\right)}}}
    \begin{tabular}{c c c c c c}
    \toprule
    \midrule
    \multicolumn{1}{l}{$\mathbf{S_3}$} & \\
        $T_1 = \SmallMatrix{
            1 & 0 & 0 \\
            0 & 1 & 0 \\
            0 & 0 & 1
        }$, &
        $T_2 = \pm \SmallMatrix{
            0 & 1 & 0 \\
            1 & 0 & 0 \\
            0 & 0 & 1
        }$, &
        $T_3 = \SmallMatrix{
            0 & 0 & 1 \\
            1 & 0 & 0 \\
            0 & 1 & 0
        }$, &
        $T_4 = \pm \SmallMatrix{
            0 & 0 & 1 \\
            0 & 1 & 0 \\
            1 & 0 & 0
        }$, &
        $T_5 = \SmallMatrix{
            0 & 1 & 0 \\
            0 & 0 & 1 \\
            1 & 0 & 0
        }$, &
        $T_6 = \pm \SmallMatrix{
            1 & 0 & 0 \\
            0 & 0 & 1 \\
            0 & 1 & 0
        }$
   \\[5pt]
    \midrule
    \multicolumn{1}{l}{$\mathbf{Mod \ S_3 \times Z_3}$} & \\
        $T_1^A = \SmallMatrix{
            1 & 0 & 0 \\
            0 & 1 & 0 \\
            0 & 0 & 1
        }$, &
        $T_2^A = \pm \SmallMatrix{
            0 & 1 & 0 \\
            1 & 0 & 0 \\
            0 & 0 & -a
        }$, &
        $T_3^A = \SmallMatrix{
            0 & 0 & 1 \\
            -1/2 & 0 & 0 \\
            0 & -1/2 & 0
        }$, &
        $T_4^A = \pm \SmallMatrix{
            0 & 0 & -a \\
            0 & a-1 & 0 \\
            -a & 0 & 0
        }$, &
        $T_5^A = \SmallMatrix{
            0 & -1/2 & 0 \\
            0 & 0 & -1/2 \\
            1 & 0 & 0
        }$, &
        $T_6^A = \pm \SmallMatrix{
            1 & 0 & 0 \\
            0 & 0 & a-1 \\
            0 & a-1 & 0
        }$
        \\ [10pt]
         $T_1^B = \SmallMatrix{
            -1/2 & 0 & 0 \\
            0 & -1/2 & 0 \\
            0 & 0 & -1/2
        }$, &
        $T_2^B = \pm \SmallMatrix{
            0 & a-1 & 0 \\
            a-1 & 0 & 0 \\
            0 & 0 & 1
        }$, &
        $T_3^B = \SmallMatrix{
            0 & 0 & -1/2 \\
            -1/2 & 0 & 0 \\
            0 & 1 & 0
        }$, &
        $T_4^B = \pm \SmallMatrix{
            0 & 0 & 1 \\
            0 & -a & 0 \\
            1 & 0 & 0
        }$, &
        $T_5^B = \SmallMatrix{
            0 & 1 & 0 \\
            0 & 0 & -1/2 \\
            -1/2 & 0 & 0
        }$, &
        $T_6^B = \pm \SmallMatrix{
            a-1 & 0 & 0 \\
            0 & 0 & -a \\
            0 & -a & 0
        }$
    \\ [10pt]
     $T_1^C = \SmallMatrix{
            -1/2 & 0 & 0 \\
            0 & -1/2 & 0 \\
            0 & 0 & -1/2
        }$, &
        $T_2^C = \pm \SmallMatrix{
            0 & -a & 0 \\
            -a & 0 & 0 \\
            0 & 0 & a-1
        }$, &
        $T_3^C = \SmallMatrix{
            0 & 0 & -1/2 \\
            1 & 0 & 0 \\
            0 & -1/2 & 0
        }$, &
        $T_4^C = \pm \SmallMatrix{
            0 & 0 & a-1 \\
            0 & 1 & 0 \\
            a-1 & 0 & 0
        }$, &
        $T_5^C = \SmallMatrix{
            0 & -1/2 & 0 \\
            0 & 0 & 1 \\
            -1/2 & 0 & 0
        }$, &
        $T_6^C = \pm \SmallMatrix{
            -a & 0 & 0 \\
            0 & 0 & 1 \\
            0 & 1 & 0
        }$    \\[5pt]
  \midrule
  \bottomrule
\end{tabular}
\caption{Ordering matrices in the $S_3$ and modulated $S_3 \times Z_3$ magnetizations.
``$+$'' (``$-$'') corresponds to the FM (AFM) $S_3$ order and the modulated $(S_3 \times Z_3)_{1 (2)}$ order.
$a \in [0,1]$ is $\abs{\Gamma/K}$ dependent.
The $S_3$ matrices form the symmetric group $S_3$.
The $S_3 \times Z_3$ matrices consist of three distinct $S_3$ sectors, featuring  a spin-lattice entangled modulation
$T^A_k + T^B_k + T^C_k = 0$.
A global sign difference is in $T_k$ with $k = 2,4,6$, reflecting the sublattice symmetry of the Hamiltonian Eq.~\eqref{eq:Hamiltonian} in zero field.
}\label{tab:ops}
\end{table*}

A standard protocol to devise spin liquids is to introduce competing orders.
In contrast to this familiar scenario, the emergence of the $S_3$ and the modulated $S_3 \times Z_3$ orders are caused here by \emph{cooperation} and \emph{competition} between two spin liquids.

\emph{Unfrustrated $S_3$ orders.}
We first discuss the two $S_3$ phases in the unfrustrated regions $K \Gamma > 0$.
The discussion will also facilitate the understanding of the more exotic $S_3 \times Z_3$ phases.

From the learned $C_{\mu\nu}$ matrices (see Appendix~\ref{app:op}), we identify that both $S_3$ orders have six magnetic sublattices with an order parameter
\begin{align} \label{eq:S3_op}
    \overrightarrow{M}_{S_3} = \frac{1}{6} \sum_{k=1}^6 T_k \vec{S}_k,
\end{align}
where $T_k$ are ordering matrices, given in Table~\ref{tab:ops},
and the FM and AFM $S_3$ order differ by a global sign in $T_2$, $T_4$, and $T_6$.
The six ordering matrices form the symmetric group $S_3$.
Its cyclic subgroup, $C_3 := \{T_1, T_3, T_5 \}$, are three-fold rotations about the $[111]$ direction in spin space, while $T_2, T_4$ and $T_6$ correspond to reflection planes $(110), (011), (101)$, respectively.
These matrices also reproduce the dual transformations that uncover the hidden $O(3)$ points residing at $K = \Gamma$ in the unfrustrated regions of the $K$-$\Gamma$ model~\cite{Chaloupka15}.

The two $S_3$ orders feature the same static spin-structure factor (SSF).
Both develop magnetic Bragg peaks at the K points of the honeycomb Brillouin zone (Figure~\ref{fig:SSF}), as the well-known $\sqrt{3} \times \sqrt{3}$ order.
This highlights the importance of knowing explicit order parameters, as different phases may display identical features in momentum space.

Furthermore, we identify the other two GSCs,
\begin{align}\label{eq:S3_GSC}
    G_{S_3} = \frac{1}{15} \corr{ \pm G_1 \pm G_2 + G_3 + G_4 \pm G_5 \pm G_6}_{\varhexagon} = 1, 
\end{align}
which equally comprise $G_{\rm KSL}$ and $G_{\rm \Gamma SL}$ in Eqs.~\eqref{eq:KSL}-\eqref{eq:GSL}, with additional $G_4$ and $G_6$ terms owing to the normalization $\abs{\vec{S}} = 1$.

As we measure in Figure~\ref{fig:measure} (a), (c), in the spin-liquid limits
$\theta \in \{0, \frac{\pi}{2}, \pi, \frac{3\pi}{2} \}$,
Kitaev and $\Gamma$ GSCs satisfy, as
$G_{\rm KSL} = \pm 1$ or $G_{\rm \Gamma SL} = \pm 1$ with other correlations vanishing.
However, when both $K$ and $\Gamma$ interactions are present and of the same sign, the two characteristic correlations $G_{\rm KSL}$ and $G_{\rm \Gamma SL}$ will lock together.
This eliminates the local symmetries of Kitaev and $\Gamma$ spin liquids and gives way to the $S_3$ orders.

It is worth noting that the two $S_3$ phases also represent rare instances where magnetic states possess non-trivial GSCs, which normally exist in cases of classical spin liquids and multipolar orders~\cite{Greitemann19b}.

\emph{Mod $S_3 \times Z_3$ phases.}
The modulated $S_3 \times Z_3$ orders can be measured by the order parameter
\begin{align} \label{eq:S3Z3_op}
    \overrightarrow{M}_{S_3 \times Z_3} = \frac{1}{18} \sum_{\alpha}^{\scalebox{0.5}{A,B,C}} \sum_{k=1}^6 T_{k}^\alpha \vec{S}_{k}^\alpha,
\end{align}
where $T_{k}^\alpha$ are eighteen ordering matrices given in Table~\ref{tab:ops}, and $\alpha = A, B, C$ distinguish three different $S_3$ sectors as illustrated in Figure~\ref{fig:PD} (b).
The $(S_3 \times Z_3)_1$ and $(S_3 \times Z_3)_2$ order differ by a global sign for all even $k$'s.

These orders exhibit a delicate \emph{spin-lattice entangled} modulation,
\begin{align} \label{eq:modulation}
    T^A_k + T^B_k + T^C_k = 0.
\end{align}
In concrete terms, $T^\alpha_{3,5}$ remain three-fold rotations along the $[111]$ direction, but there is an additional $\cos(2\pi/3)$ factor entering some, but not all, spin components.
The location of this factor, as shown in Table~\ref{tab:ops}, alternates among the three $S_3$ sectors, to achieve the cancellation in Eq.~\eqref{eq:modulation}.
Furthermore, mirror reflections, $T^\alpha_k$ with even $k$'s are decorated by a factor $a \in [0, 1]$, in such a way that a cancellation with the mirror of the same type occurs, as $(a-1) + (-a) + (1) = 0$.
The value of $a$, which TK-SVM also identifies, strongly depends on the relative strength $\abs{\Gamma/K}$, while the reflection planes remain locked on $(110), (011), (101)$.

This modulation is very different from those in multiple-$\mathbf{q}$ orders and spin-density-wave (SDW) orders where phase factors universally act on all spin components.
Moreover, since this modulation does not preserve spin length, the $S_3 \times Z_3$ magnetization will not saturate to unity, but to a reduced value $M \lesssim \frac{2}{3}$, reflecting an intrinsic frustration.

The SSF of the two $S_3 \times Z_3$ phases is shown in Figure~\ref{fig:SSF}~(b). The large magnetic cell leads to a reduced Brillouin zone. 
The SSF pattern nevertheless only partially reveals properties of the ordering and does not show information of the spin-lattice entangled modulation in Eq.~\eqref{eq:modulation}, again underlining the significance of analytical order parameters.

To better understand the nature of the modulated $S_3 \times Z_3$ orders, we show their magnetization along with the $G_{\rm KSL}$ and $G_{\rm \Gamma SL}$ correlations in Figure~\ref{fig:measure} (b) and (d).
To exclude the $\abs{K/\Gamma}$-dependence in the order parameter, we defined an alternative magnetization by including only odd $k$'s in Eq.~\eqref{eq:S3Z3_op},
$\protect\overrightarrow{M}^{\prime}_{S_3 \times Z_3} = \frac{1}{9} \sum_{\alpha}^{\scalebox{0.5}{A,B,C}} \sum_{k}^{\scalebox{0.6}{1,3,5}} T_{k}^\alpha \vec{S}_{k}^\alpha$.
Clearly, in the frustrated regions, the characteristic Kitaev and $\Gamma$ correlations develop toward opposite directions.
Near the Kitaev limits, $\theta = \frac{\pi}{2}, \frac{3\pi}{2}$, $G_{\rm KSL}$ dominates;
the system stays disordered, either in an extended KSL phase or a CP region.
When $G_{\rm \Gamma SL}$ is sufficiently strong to compete with $G_{\rm KSL}$, at $\abs{\Gamma/K} \sim 0.27$,
an $S_3 \times Z_3$ order emerges from the two conflicting quadratic correlations, and expands till the large $\Gamma$ limits owing to the small exGSD of a $\Gamma$SL.

Because of the relevance to the spin-liquid candidate $\alpha$-$\mathrm{RuCl}_3$, (a part of) the parameter regime with FM $K$ and intermediate AFM $\Gamma$ has attracted much attention, as the $\Gamma$ term in this material is found to be comparable to the Kitaev interaction~\cite{Ran17, Kim16, Yadav16, Winter16}.
On the one hand, exact diagonalization (ED) of small systems~\cite{Gordon19}, infinite density matrix renormalization group [(i)DMRG] simulations on narrow cylinders~\cite{Gordon19, Jiang19, Gohlke20}, classical Luttinger-Tisza~\cite{Rau14}, and cluster mean-field~\cite{Rusnacko19} analyses observed there a disordered phase or incommensurate order.
On the other hand, classical simulated-annealing calculations for small system sizes~\cite{Chern20} and simple-update infinite projected-entangled-pair state (iPEPS) simulations~\cite{Lee20} reported magnetic states with enlarged unit cells but of unknown nature. Our results are compatible with the latter observations.
The magnetic Bragg peaks (located at the $\frac{2}{3}{\bf M}$ points) of the $(S_3 \times Z_3)_2$ phase are consistent with the SSFs reported in Ref.~\cite{Chern20}. However, our machine identifies the order parameter and the correlations underlying the phase.

The fate of the modulated $S_3 \times Z_3$ order in quantum $K$-$\Gamma$ models, for the case of spin-$1/2$ as well as higher $S$ values, is open and left for future studies.
It is however not uncommon that, when a system establishes a robust magnetic order in the classical large-$S$ limit, this order can  persist in the quantum cases with a reduced ordering moment due to quantum fluctuations. Such examples are known for various spin-liquid candidates, see for instance Refs.~\cite{Zheng06, White07, Li18}. 

The firmness of the $S_3 \times Z_3$ orders can be demonstrated in several ways.
In Figure~\ref{fig:measure} (b) and (d), we confirm their stability by varying $\theta = \arctan (K/\Gamma) $ over the entire frustrated region.
Moreover, the global phase diagram Figure~\ref{fig:PD} (c) shows that they are robust against finite fields.
This is further verified in Appendix~\ref{app:S3Z3} where we show a direct Monte Carlo measurement of the $(S_3 \times Z_3)_2$ order and its suppression in intermediate fields.
In addition, the stability of this order against thermal fluctuations, inevitable for real systems, is also established in Appendix~\ref{app:S3Z3}.
Interestingly, the melting involves two stages and gives rise to an intermediate paramagnetic regime found for temperatures significantly below the Neel temperature.

From the machine learning point of view, the two modulated $S_3 \times Z_3$ orders provide a hallmark of a machine-learning algorithm identifying different, complicated phases.
Furthermore, the identification of the spin-liquid constraints also gives insight into their origin, by which the emergence of magnetic orders in the $K$-$\Gamma$ model can be consistently explained.

\section{Conclusions}\label{sec:summary}
Machine learning techniques are emerging as promising tools in various disciplines of physics~\cite{Carleo19}.
However,  results going beyond the state of the art are required before they will disrupt current procedures.
By subjecting the honeycomb $K$-$\Gamma$-$h$ model to the analysis of our unsupervised and interpretable TK-SVM method, we have shown that machine learning can indeed handle highly complicated problems in  frustrated magnets and reveal unknown physics. 

We found that the classical phase diagram of the $K$-$\Gamma$ model in an $[111]$ field is exceptionally rich (see Figure~\ref{fig:PD}), hosting several unconventional symmetry-breaking phases and a plethora of disordered states at very low temperature.
The phase diagram clearly shows the finite extent of the KSLs, an intermediate disordered phase at the AFM Kitaev limit, and a field-induced suppression of magnetic orders, which were previously only reported for quantum systems.
These common features strongly suggest that certain aspects of the Kitaev materials can already be understood from a semi-quantitative classical picture and also call for a systematic investigation of larger spin models in order to find  potential higher-$S$ spin liquids.

Two phases, the modulated $S_3 \times Z_3$ magnets, with a previously unknown type of modulation were identified.
On the one hand, these states represent a concrete instance of machine learning successfully discovering novel phases.
Their structure is sufficiently complicated, but it is picked up without difficulty by TK-SVM.
On the other hand, they also imply that the competition between Kitaev and non-Kitaev exchanges can significantly enrich the physics and lead to more unconventional phases than expected.
 
We discovered the GSCs of the classical $\Gamma$SLs and reproduced the ones of the KSLs.
Not only did these constraints enhance our understanding of the $\Gamma$SLs, they also put the emergence of the complicated orders in the $K$-$\Gamma$ model in a \emph{unifying} picture.
The two unfrustrated $S_3$ magnets emerge when the characteristic Kitaev and $\Gamma$ correlations cooperatively eliminate the macroscopic degeneracy of each other.
By contrast, the two modulated $S_3 \times Z_3$ magnets can be understood as the consequence of the competition between the KSL and $\Gamma$SL.
This mechanism may be viewed as an alternative protocol for devising exotic phases.

Our work may stimulate future applications of machine learning in Kitaev materials and beyond.
The study of Kitaev materials is motivated by realizing the Kitaev model~\cite{Jackeli09, Chaloupka10}.
In real systems, non-Kitaev interactions are ubiquitously present and cannot be treated as perturbations.
In the case of $\alpha$-$\mathrm{RuCl}_3$, aside from the dominating Kitaev and $\Gamma$ exchanges, the Heisenberg $J_1$, $J_3$, and, possibly, the off-diagonal $\Gamma^{\prime}$ term also play a role~\cite{Laurell20, Maksimov20}.
Temperature and external fields add further dimensions to the physical parameter space~\cite{Kasahara18, Banerjee17, Bachus20}.
Similar complications are also encountered in other candidate compounds such as ${\rm A}_2{\rm IrO}_3$ ($\rm{A} = {\rm Na, K}$)~\cite{Yadav18, Yadav19} and the three-dimensional hyper- and stripy-honeycomb materials $\beta$-, $\gamma$-${\rm Li}_2{\rm IrO}_3$~\cite{Modic14, Takayama15, Biffin14}.
While these additional terms besides the Kitaev exchange can enrich the underlying physics, they also dramatically complicate the analysis.
Machine learning is designed to discover complex structures in high-dimensional data.
In the framework of TK-SVM, partitioning a phase diagram can be formulated as a two-dimensional Laplacian matrix~\cite{Fiedler73, Fiedler75}, independent of the number of physical parameters.
This ability permits an efficient scanning over complex, multi-dimensional phase diagrams. The nature of each phase will also be uncovered in virtue of the machine's interpretability.  TK-SVM may hence speed up our understanding of competing interactions in a multi-dimensional parameter space, which can in turn facilitate the experimental search and theoretical development for exotic phases.

\section*{Open source and data availability}
The TK-SVM library has been made openly available with documentation and examples~\cite{Jonas}.
The data used in this work are available upon request.

\begin{acknowledgements}
We wish to thank Hong-Hao Tu, Simon Trebst, Nic Shannon and Stephen Nagler for helpful discussions.
KL, NS, NR, JG, and LP acknowledge support from FP7/ERC Consolidator Grant QSIMCORR, No. 771891, and the Deutsche Forschungsgemeinschaft (DFG, German Research Foundation) under Germany's Excellence Strategy -- EXC-2111 -- 390814868.
Our simulations make use of the $\nu$-SVM formulation~\cite{Scholkopf00}, the LIBSVM library~\cite{Chang01, Chang11}, and the ALPS\-Core library~\cite{Gaenko17}.
\end{acknowledgements}

\begin{appendix}

\section{Setting up of TK-SVM } \label{app:op}

\begin{figure}[t]
\centering
     \begin{tabular}{c}
          \subfloat[FM $S_3$]{\centering\includegraphics[width=0.45\textwidth]{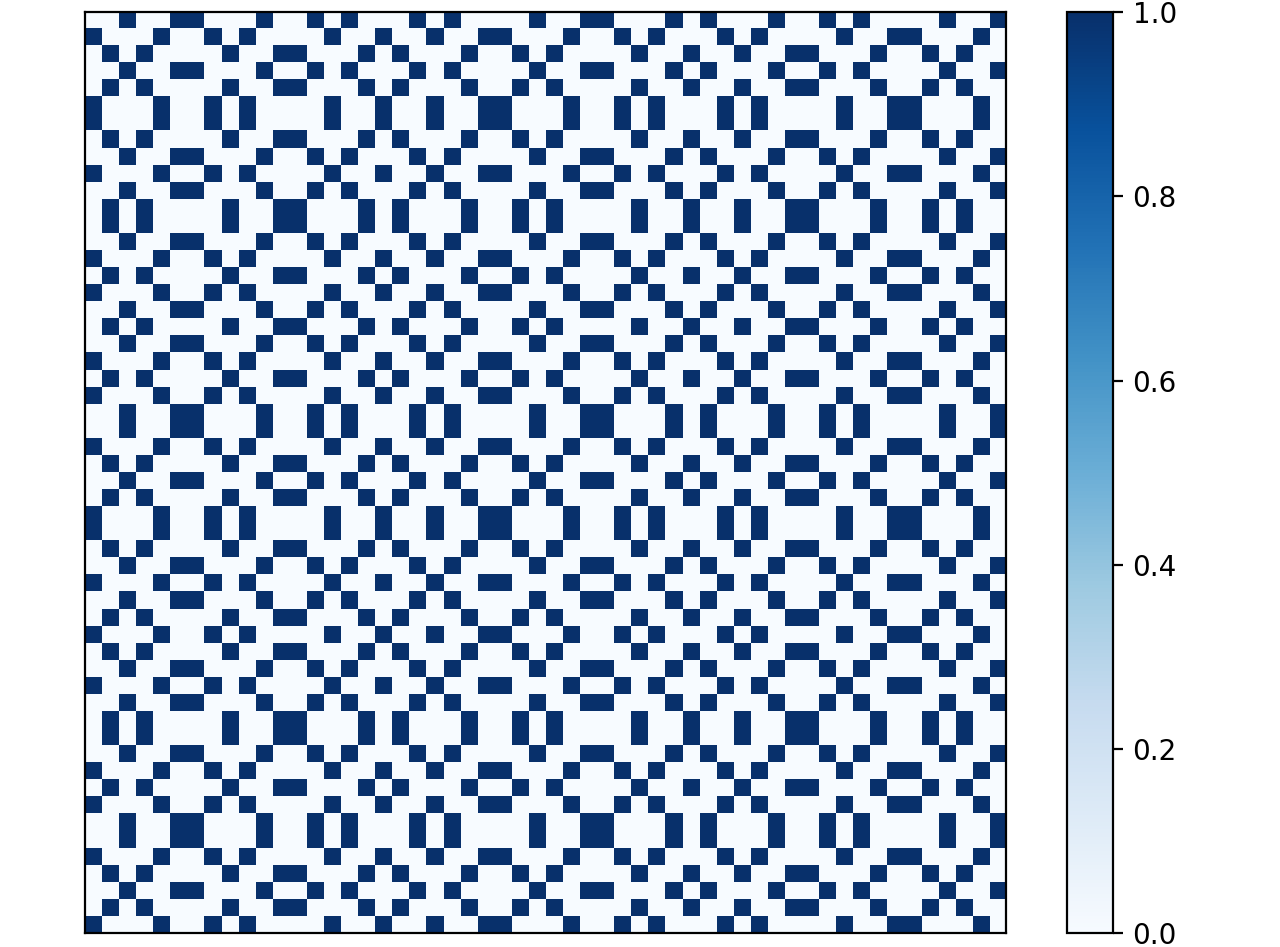}} \\
          \subfloat[mod $(S_3 \times Z_3)_2$]{\centering\includegraphics[width=0.45\textwidth]{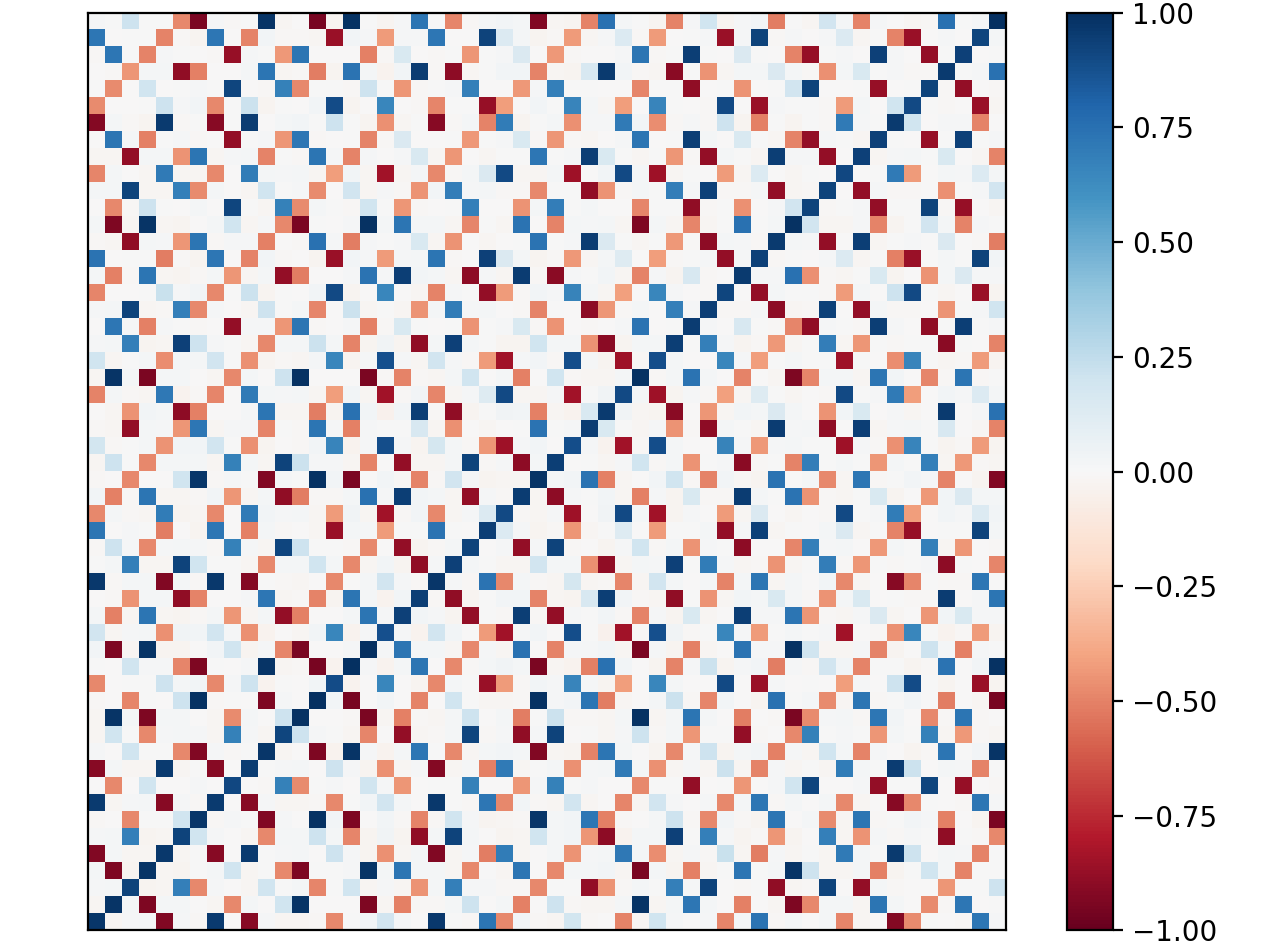}}
     \end{tabular}
 \caption{Visualization of the $C_{\mu\nu}$ matrix of the FM $S_3$ and the mod $(S_3 \times Z_3)_2$ phase. Each pixel corresponds to an entry of $C_{\mu\nu}$. Non-vanishing entries identify the relevant components of $\phi_\mu$ entering the order parameter. Here results of a $18$-spin cluster are shown for demonstration, while much larger clusters are used for the phase diagram Figure~\ref{fig:PD}. The $S_3$ order is represented multiple times as its magnetic cell has six sublattices.}\label{fig:C_pattern}
 \end{figure}

The TK-SVM method has been introduced in our previous work~\cite{Greitemann19, Liu19, Greitemann19b}. Here we review its essential ingredients for completeness.

For a sample $\mathbf{x} = \{S_i^a | i = 1, 2, ..., N; a = x,y,z \}$,
the feature vector $\bds{\phi} = \{\phi_\mu\}$ maps $\mathbf{x}$ to degree-$n$ monomials
\begin{align}\label{eq:phi}
    \phi_{\mu} = \corr{S_{\alpha_1}^{a_1} S_{\alpha_2}^{a_2} \dots S_{\alpha_n}^{a_n}}_{\rm cl},
\end{align}
where $\corr{\cdots}_{\rm cl}$ represents a lattice average up to a cluster of $r$ spins;
$\alpha_1, \dots, \alpha_n$ label spins in the cluster;
$\mu = \{\alpha_1, a_1; \dots, \alpha_n, a_n \}$ are collective indices.

TK-SVM constructs from $\phi_{\mu}$ a tensorial feature space ($\bds{\phi}$-space) to host potential orders~\cite{Greitemann19, Liu19}.
The capacity of the $\bds{\phi}$-space depends on the degree ($n$) of monomials and the size ($r$) of the cluster.
As the minimal $n$ and $r$ are unknown parameters, in practice, we choose large clusters according to the Bravais lattice and $n \in~[1, 6]$, where $n = 1$ detects magnetic orders and $n > 1$ probes multipolar orders and emergent local constraints.
In learning the phase diagram Figure~\ref{fig:PD}, we constructed $\bds{\phi}$-spaces using clusters up to $288$ spins ($12 \times 12$ honeycomb unit-cells) at rank-$1$ and clusters up to $18$ spins at rank-$2$, much beyond the needed capacity.
We also confirmed the results are consistent when varying the size and shape of clusters and found ranks $n \geq 3$ to be irrelevant.

The coefficient matrix $\mathbbm{C} = \{C_{\mu\nu}\}$ measures correlations of $\phi_{\mu}$, defined as
\begin{align}\label{eq:C_munu}
    C_{\mu \nu} = \sum_k \lambda_k \phi_\mu(\mathbf{x}^{(k)}) \phi_\nu(\mathbf{x}^{(k)}),
\end{align}
where the Lagrange multiplier $\lambda_k$ denotes the weight of the $k$-th sample and is solved in the underlying SVM optimization problem~\cite{Greitemann19, Liu19}.
Its non-vanishing entries identify the relevant basis tensors of the $\bds{\phi}$-space, and their interpretation leads to order parameters.

In Figure~\ref{fig:C_pattern}, we show the $C_{\mu\nu}$ matrix of the FM $S_3$ and the mod $(S_3 \times Z_3)_2$ phase for example.
The corresponding order parameters are given in Eqs.~\eqref{eq:S3_op} and~\eqref{eq:S3Z3_op} and are measured in Figure~\ref{fig:measure} in the main text.

\section{Details of Graph Partitioning} \label{app:graph}

\begin{figure}[t!]
  \centering
  \includegraphics[width=0.45\textwidth]{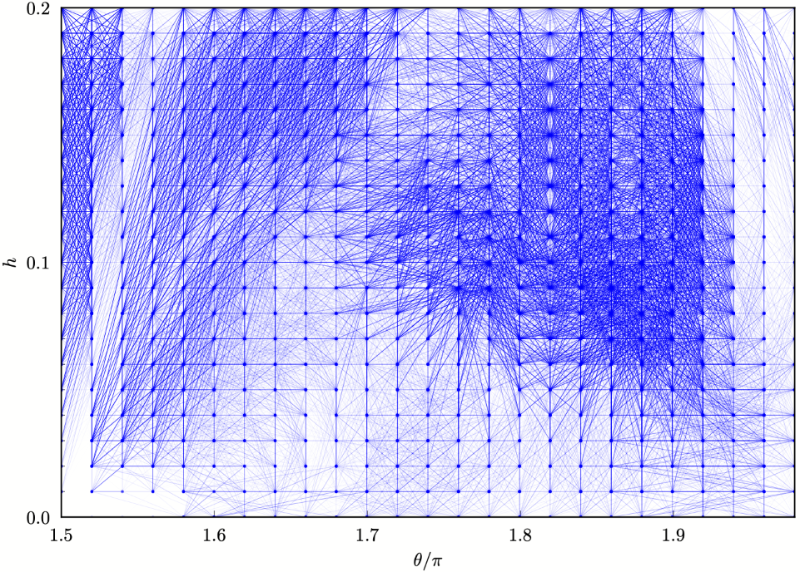}
  \caption{The $\theta \in [\frac{3\pi}{2}, 2\pi)$ sector
   of the graph is shown for visualization.
  Each vertex labels a $(\theta, h)$ point, following a uniform distribution $\Delta \theta = 0.02\pi$, $\Delta h = 0.01$.
  The edges connecting two vertices are determined by $\rho$ in the corresponding decision function and the weight function Eq.~\eqref{eq:weight}.
  Edge weights are weakened to reduce visual density.
  The entire graph contains $M = 1,\!250$ vertices with $\theta \in [0, 2\pi)$ and $M(M-1)/2 = 780,\!625$ edges, whose partition gives the phase diagram Figure~\ref{fig:PD} (c).}
  \label{fig:graph}
\end{figure}

\begin{figure}[t!]
\centering
\includegraphics[width=0.45\textwidth]{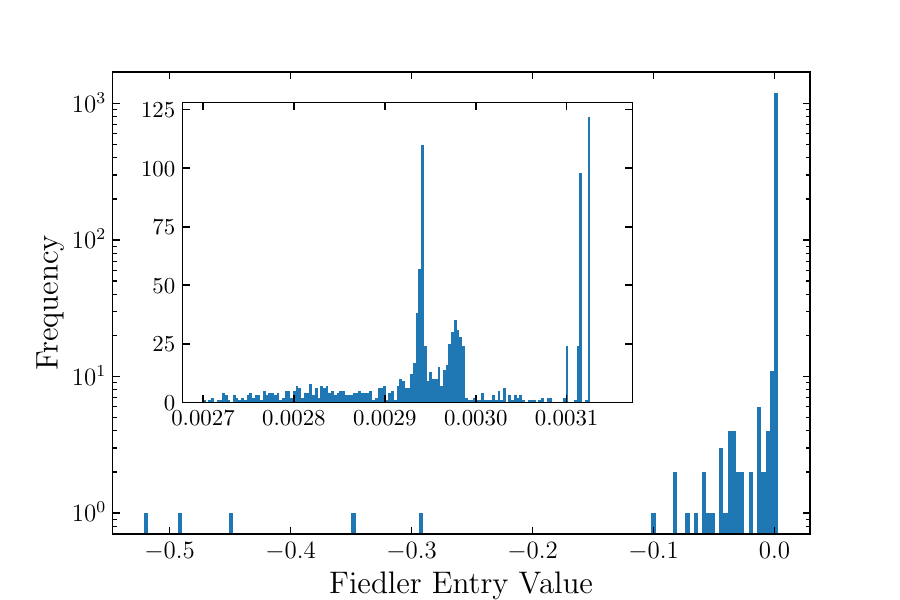}
 \caption{Histogram of Fiedler vector entries.
 Each entry corresponds to a vertex of the graph, namely, a $(\theta,h)$ point.
 Their values are color-coded by the phase diagram Figure~\ref{fig:PD} (c).
 A logarithmic scale is used in the main panel as the histogram is spanning several orders. 
 The inner panel uses a linear scale and shows a zoom-in view of the bulk of the distribution.
 From left to right, the five profound peaks in the inner panel correspond to the two $S_3 \times Z_3$ phases, the FM $S_3$, the AFM $S_3$ phase and the full polarized phase, respectively.
 Flat regions correspond to correlated paramagnets and indicate wide crossovers to neighboring phases.
}
 \label{fig:histogram}
 \end{figure}

Not all $C_{\mu\nu}$ matrices need to be interpreted. In the graph partitioning, where the goal is to learn the topology of the phase diagram, it suffices to analyze the bias parameter $\rho$.
When ${\rm A, B}$ are two phase points where spin configurations are generated, the bias parameter $\rho_{\rm AB}$ in the corresponding binary classification problem behaves as
\begin{align}\label{eq:rho_rules}
    \abs{\rho_{\rm AB}} \begin{cases}
        \gg 1 & \textup{${\rm A, B}$ in the same phase}, \\
        \lesssim 1 & \textup{${\rm A, B}$ in different phases}.
    \end{cases}
\end{align}
Thus, as demonstrated in our previous work, $\rho$ can detect phase transitions and crossovers~\cite{Liu19, Greitemann19b}.
(Though the sign of $\rho_{\rm AB}$ also has physical meaning and can reveal which phase is in the (dis-)ordered side, the absolute value is sufficient for the graph partitioning; see Ref.~\onlinecite{Greitemann19b} for details.)

The graph partitioning in TK-SVM is a systematic application of the $\rho$ criteria Eq.~\eqref{eq:rho_rules}.
The graph is built from $M = 1,\!250$ vertices, each corresponding to a point $(\theta, h)$, and $M(M-1)/2$ connecting edges; as exemplified in Figure~\ref{fig:graph}.
The weight of an edge is defined by $\rho$ in the SVM classification between the two endpoints, with a Lorentzian weighting function
\begin{align}\label{eq:weight}
  w(\rho) &= 1 - \frac{\rho_c^2}{(|\rho|-1)^2+\rho_c^2} \in [0,1].
\end{align}
Here $\rho_c$ sets a characteristic scale for ``$\gg 1$'' in Eq.~\eqref{eq:rho_rules}, as a larger $\rho_c$ tends to suppress weight of the edges.
The choice of $\rho_c$ is not critical since points in the same phase are always more connected than those from different phases.
In computing the phase diagram Figure~\ref{fig:PD}, $\rho_c = 1000$ is applied, but we also verified that the results are robust when $\rho_c$ is changed over an interval ranging from a small $\rho_c = 10$ to a large $\rho_c = 10^4$, where all edge weights are almost eliminated.

A graph with $10^6$ edges is considered  a small problem in graph theory and may be partitioned with different methods.
We have applied Fiedler's theory of spectral clustering~\cite{Fiedler73, Fiedler75}.
The result is a so-called Fiedler vector of the dimensionality $M$, corresponding to the $M$ vertices.
Strongly connected vertices, namely those in the same phase, share equal or very close Fiedler-entry values, while those in different phases have substantially different Fiedler entries.
In this sense, the Fiedler vector can act as a phase diagram.

\begin{figure}[t]
  \centering
  \includegraphics[width=0.44\textwidth]{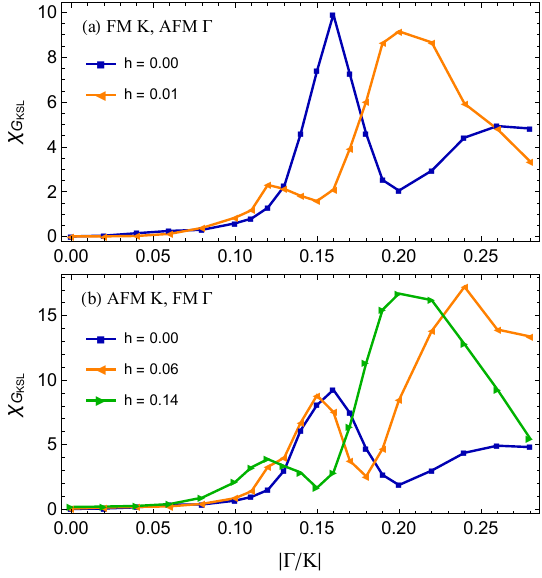}
  \caption{Susceptibility for the characteristic Kitaev correlation $G_{\rm KSL}$ as function of $\abs{\Gamma/K}$, in the vicinity of the FM~(a) and AFM~(b) Kitaev limit with $K\Gamma \leq 0$.
  The first peak of $\chi_{\rm G_{KSL}}$ in a fixed $h$ identifies the crossover from a classical KSL to a non-Kitaev correlated paramagnet.
  At $h=0$, the KSLs survive until $\abs{\Gamma/K} \sim 0.16$.
  When magnetic fields are applied, the peak moves consistently towards a smaller value of $\abs{\Gamma/K}$ with its width broadening.
  The wide bumps at lager $\abs{\Gamma/K}$ signal the second crossover to a modulated $S_3\times Z_3$ phase, for which the optimal quantity is the $S_3\times Z_3$ magnetization.}
  \label{fig:chi}
\end{figure}

Note that, the two-dimensional representation of the graph shown in Figure~\ref{fig:graph} is only for visualization.
Regardless of the dimension of a physical parameter space, a graph can always be formulated by a Laplacian matrix, and its partitioning gives a vectorial quantity, i.e., the Fiedler vector~\cite{Fiedler73, Fiedler75, Liu19}.

Figure~\ref{fig:histogram} shows the histogram of the Fiedler entries for the phase diagram Figure~\ref{fig:PD} (c), which clearly exhibit a multinodal structure.
Each peak corresponds to a distinct phase, and the wide bumps are indicative of crossover regions or phase boundaries.

\begin{figure}[t]
  \centering
  \includegraphics[width=0.425\textwidth]{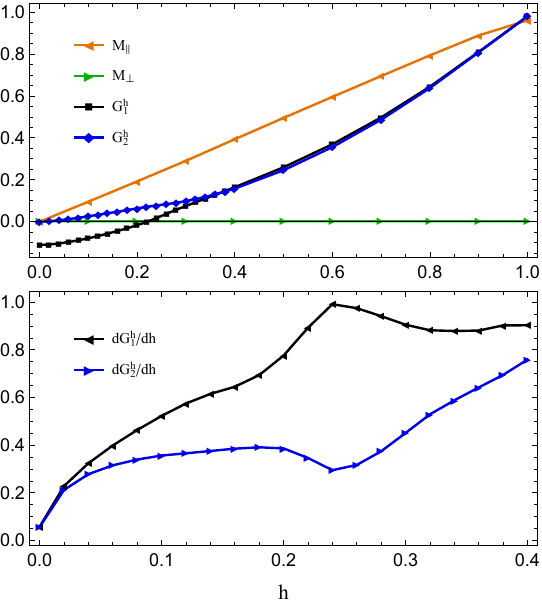}
  \caption{Field dependence of the  magnetization per spin parallel ($m_{\parallel}$) and perpendicular ($m_{\perp}$) to the $[111]$ field, and the normalized $U(1)_g$-symmetric correlations, $G^h_{1}$ and $G^h_{2}$, at the AFM Kitaev limit $(K, \Gamma) = (1, 0)$.
  The spins are mostly paramagnetic under weak and intermediate fields. Bumps in $dG^h/dh$ may imply prominent changes in the system.}
  \label{fig:Gh}
\end{figure}

\begin{figure}[t]
  \centering
  \includegraphics[width=0.47\textwidth]{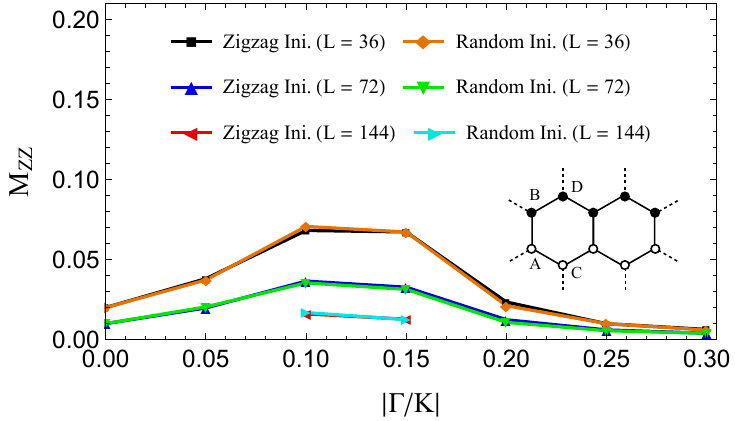}
  \caption{Monte Carlo measurement of the zigzag order in the region of FM $K$ and small AFM $\Gamma$ at $h = 0$, where the magnetization
  $M_{\rm ZZ} = \big\langle\big|\frac{1}{N_{\rm cell}}\sum_{\rm cell} \big(\vec{S}_A - \vec{S}_B + \vec{S}_C - \vec{S}_D \big) \big|\big\rangle$, and $A, B, C, D$ label the four sub-lattices.
Simulations initiated with perfect zigzag states are compared with random initializations.
The zigzag order appears to be unstable in all cases.
The small residual moments are a finite-size effect and decrease significantly with increasing system sizes.}
  \label{fig:ZZ}
\end{figure}

\section{Extension of Classical KSLs}\label{app:measure}

Since a GSC, $G$, characterizes a classical spin liquid, we can accordingly define a susceptibility to measure how sharp it is defined,
\begin{align}\label{eq:chi}
\chi_{\rm G} = \frac{V}{T}\left( \corr{G^2} - \corr{G}^2\right),
\end{align}
where $\corr{\dots}$ is the ensemble average, and $V$ denotes the volume of the system.
Such a susceptibility was first introduced in Ref.~\cite{Greitemann19b}, and we showed with various examples its high sensitivity to the breakdown of an associated classical spin liquid.

To estimate the extension of classical KSLs, we define  the susceptibility $\chi_{\rm G_{KSL}}$. It is shown in Figure~\ref{fig:chi} as a function of the competing $\Gamma$ interaction.
At a fixed $h$, $\chi_{\rm G_{KSL}}$ develops two peaks/bumps, reflecting the violation of the GSC.
The sharper peak at a smaller $\abs{\Gamma/K}$ is responsible for the crossover between a KSL and a non-Kitaev correlated paramagnet.
The broad bump at a larger $\abs{\Gamma/K}$ signals the second crossover to a modulated $S_3 \times Z_3$ phase.
(The optimal measure to this second crossover is the $S_3 \times Z_3$ order parameter instead of $\chi_{\rm G_{KSL}}$. However, the location of the bump qualitatively agrees with the results based on the $S_3 \times Z_3$ magnetization, see Figure~\ref{fig:measure} for example.)

In order to examine the effects of magnetic fields on the AFM KSL, we measure the field-dependence of the two $U(1)_g$-symmetric correlations, $G^h_1$ and $G^h_2$, and the  magnetization per spin parallel ($m_{\parallel}$) and perpendicular ($m_{\perp}$) to the $[111]$ field, as is shown in Figure~\ref{fig:Gh}.
Under weak and intermediate fields, most of the spins respond paramagnetically, as $m_{\parallel}$ is small and $m_{\perp}$ vanishes.
While $G^h_1$ and $G^h_2$ smoothly increase with the external field, the bumps in their derivative may imply prominent changes in the system, which are used to estimate the extent of the AFM KSL.
The regime with intermediate field is marked as a $U(1)_g$ region in order to distinguish it from a polarized state.
In the main text (see Sections~\ref{sec:phase_diagram} and~\ref{sec:sls}), we discussed that this regime coincides with a gapless spin liquid proposed for quantum spin-$1/2$ and spin-$1$ AFM Kitaev models~\cite{Motome20, Zhu18, Hickey19, Hickey20, Zhu20}.
A  similar segmentation in the finite-$h$ phase diagram is observed in the quantum case~\cite{Zhu18, Zhu20}.

The behavior of $\chi_{\rm G_{KSL}}$, $G^h_1$, and $G^h_2$ are used to estimate the boundary (indicated by the dashed lines in Figure~\ref{fig:PD} (c)) between the KSLs and other correlated paramagnets, supplementing the graph partitioning.
This is needed because, in the graph partitioning shown in Figure~\ref{fig:PD}, we only employed a rank-$1$ TK-SVM which is designed for detecting the presence and absence of magnetic order.
To classify different spin liquids, we use rank-$2$ TK-SVM to identify their GSCs.
In principle, we could also have performed a separate graph partitioning with rank-$2$ TK-SVM. But, given the rank-$1$ results, most of the phase diagram has already been fully classified this way and there are only few locations left worth examining at higher rank.

In Figure~\ref{fig:ZZ} we evaluate the zigzag magnetization in the extended KSL and CP region with FM $K$ and AFM $\Gamma$.
The zigzag order has been considered as a $\Gamma$-induced competing order to a KSL.
However, it can be shown that it is unstable at low temperature and experiences strong finite-size effects.
This is also consistent with the picture that, in order to stabilize the zigzag-like order found in $\alpha$-$\mathrm{RuCl}_3$~\cite{Banerjee17, Ran17}, other terms, such as the first and third nearest-neighbor Heisenberg $J_1$, $J_3$ interactions~\cite{Gohlke18, Gass20, Jiang19} or the off-diagonal $\Gamma^{\prime}$ term~\cite{Gordon19, Lee20, Gohlke20, Chern20}, are needed.

\begin{figure}[t]
  \centering
  \includegraphics[width=0.44\textwidth]{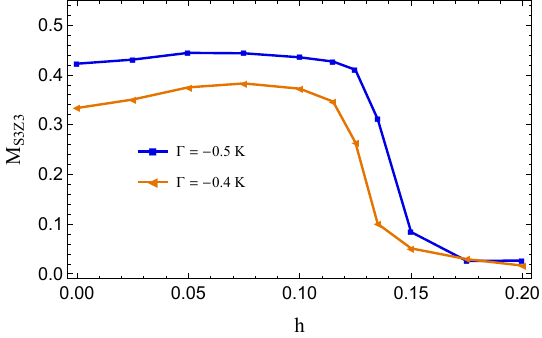}
  \caption{Monte Carlo measurement of the $(S_3 \times Z_3)_2$ magnetization as a function of the $[111]$ field, in the region of FM $K$ and intermediate AFM $\Gamma$. The $(S_3 \times Z_3)_2$ magnetization extends over a finite region of external field and is subsequently suppressed to a small but finite value, see Figure~\ref{fig:PD}(c).}
  \label{fig:S3Z3_h}
\end{figure}

\begin{figure}[t]
  \centering
  \includegraphics[width=0.44\textwidth]{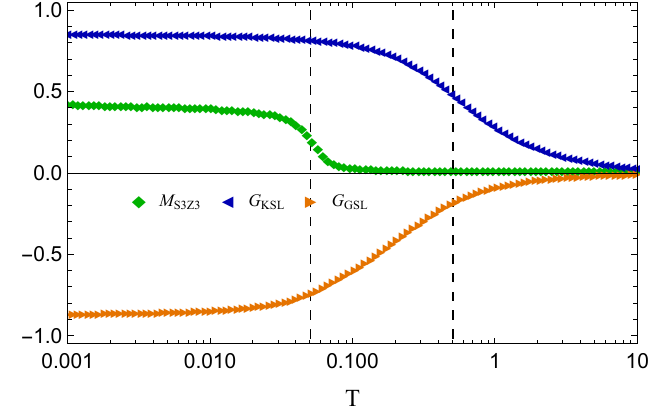}
  \caption{Temperature dependence of the $(S_3 \times Z_3)_2$ order and the corresponding Kitaev and $\Gamma$ correlations for $h = 0$, $\Gamma = -0.5$, and $ K > 0$.
The system exhibits a two-step melting, dividing the temperature range into three regimes.
In the low-temperature regime, the $(S_3 \times Z_3)_2$ order is established along with strong $G_{\rm KSL}$ and $G_{\rm \Gamma SL}$.
The intermediate regime is a correlated paramagnet, where the competing Kitaev and $\Gamma$ correlations are already noticeable but not strong enough to stabilize magnetic order.
A trivial paramagnet is found for high temperatures. The dashed lines mark the location of crossovers.}
  \label{fig:S3Z3_T}
\end{figure}

\section{Field and temperature dependence of the modulated $(S_3 \times Z_3)_2$ order}\label{app:S3Z3}

The machine-learned global phase diagram Figure~\ref{fig:PD} (c) shows that the modulated $S_3 \times Z_3$ orders extend over a finite region of a $[111]$ field.
In particular, in the parameter regime relevant for $\alpha$-$\mathrm{RuCl}_3$, namely a FM $K$ and an intermediate AFM $\Gamma$, the $(S_3 \times Z_3)_2$ order experiences a field-induced suppression.
This is further confirmed in Figure~\ref{fig:S3Z3_h} by direct measurement of the $S_3 \times Z_3$ magnetization.
After suppressing the order, the system enters a partially-polarized frustrated regime, owing to the competition between the external field and the Kitaev and $\Gamma$ interactions.
A similar classical regime was discussed in Ref.~\cite{Gohlke20}  and was considered as the parent phase of two quantum nematic paramagnets in the spin-$1/2$ $K$-$\Gamma$ model~\cite{Gohlke20, Lee20}.

In Figure~\ref{fig:S3Z3_T}, we evaluate the temperature dependence of the $(S_3 \times Z_3)_2$ magnetization and the corresponding Kitaev and $\Gamma$ correlations. The system exhibits two crossovers when increasing temperature.
Order is established in the low-temperature regime with strong $G_{\rm KSL}$ and $G_{\rm \Gamma SL}$.
Its melting is followed by an intermediate regime where the Kitaev and $\Gamma$ correlations already develop but are not yet strong enough to stabilize magnetic order.
This is consistent with the scenario discussed in Section~\ref{sec:orders} that the $S_3 \times Z_3$ order can be understood from the competition between the two quadratic correlations.
This intermediate regime also extends until nearly one order below the Neel temperature which is set by the interaction strength, and may hence be viewed as a finite-temperature correlated paramagnet. 

While a two-step melting is often observed for spin liquids, including the quantum KSL~\cite{Motome20, Nasu17} and the classical $\Gamma$SL~\cite{Saha19}, as well as for  spin nematics, such as the multipolar orders in the Kagome~\cite{Zhitomirsky08} and pyrochlore~\cite{Taillefumier17} anti-ferromagnets, such a phenomenon is quite unusual for a magnetically ordered system.
We leave for future studies to find out what type of excitations are responsible for the two crossovers, and whether such a two-step melting can also be present when other interactions that can exist in real materials are included.

The $(S_3 \times Z_3)_1$ phase has the same temperature dependence because of the sub-lattice symmetry of the $K$-$\Gamma$ model at zero field.
However, the sign of $G_{\rm KSL}$ and $G_{\rm \Gamma SL}$ is swapped as in the case of Figure~\ref{fig:measure} (b) and (d).
\end{appendix}

\bibliography{jkgm}

\end{document}